# Magneto-transport and magnetic susceptibility of SmFeAsO$_{1-x}$F$_x$ ($x$ = 0.0 and 0.20)

R. S. Meena[1,2], Anand Pal[1], Shiva Kumar[1], K. V. R Rao[2], and V.P.S. Awana[1]*

[1]Quantum Phenomena and Application Division, National Physical Laboratory (CSIR)
Dr. K. S. Krishnan Road, New Delhi-110012, India
[2]Department of Physics, University of Rajasthan, Jaipur-302055, India

Bulk polycrystalline samples, SmFeAsO and the iso-structural superconducting SmFeAsO$_{0.80}$F$_{0.20}$ are explored through resistivity with temperature under magnetic field ρ(T, H), *AC* and *DC* magnetization (*M-T*), and Specific heat ($C_p$) measurements. The Resistivity measurement shows superconductivity for $x$ = 0.20 sample with $T_c$(onset) ~ 51.7K. The upper critical field, [H$_{c2}$(0)] is estimated ~3770kOe by Ginzburg-Landau (GL) theory. Broadening of superconducting transition in magneto transport is studied through thermally activated flux flow in applied field up to 130 kOe. The flux flow activation energy (U/k$_B$) is estimated ~1215K for 1kOe field. Magnetic measurements exhibited bulk superconductivity with lower critical field ($H_{c1}$) of ~1.2kOe at 2K. In normal state, the paramagnetic nature of compound confirms no trace of magnetic impurity which orders ferromagnetically. *AC* susceptibility measurements have been carried out for SmFeAsO$_{0.80}$F$_{0.20}$ sample at various amplitude and frequencies of applied *AC* drive field. The inter-granular critical current density ($J_c$) is estimated. Specific heat [$C_p$(T)] measurement showed an anomaly at around 140K due to the *SDW* ordering of Fe, followed by another peak at 5K corresponding to the antiferromagnetic (*AFM*) ordering of Sm$^{+3}$ ions in SmFeAsO compound. Interestingly the change in entropy (marked by the $C_p$ transition height) at 5K for Sm$^{+3}$ *AFM* ordering is heavily reduced in case of superconducting SmFeAsO$_{0.80}$F$_{0.20}$ sample.

Key Words: Oxy-Arsenide superconductor family, Superconductivity, X-ray diffraction, Rietveld analysis, Magnetization and Thermally activated flux flow.

**Introduction:**

The discovery of superconductivity in layered Iron-based LaFeAsO$_{1-x}$F$_x$ compound at 26K had attracted much attention of the condensed matter researchers [1]. The finding of superconductivity in these compounds is surprising due to the presence of familiar magnetically ordered iron. The superconducting transition temperature of this compound has been increased up to 43K under a high pressure of ~4Gpa [2]. Followed by this result $T_c$ increased up to 56K through creating a chemical pressure by replacing La with other rare earth element having lower ionic radii. This is highest $T_c$ except high temperature cuprates [3-6]. Structurally, *RE*FeAsO compounds resemble the cuprate superconductors, but based on anti-PbO-type conducting FeAs layers rather than planer CuO$_2$ layer. Similar to High $T_c$ cuprates, iron pnictides are layered compounds with alternating Fe-As layers sandwiched between non conducting RE-O layers. The parent compounds of cuprates are Mott insulators with localized magnetic moments, on the other hand undoped pnictides are metallic with itinerant magnetic moments. The superconductivity in these materials does not follow the BCS theory and they are known to be unconventional superconductors. Therefore careful effort is required to elucidate the interesting physics of iron pnictide compounds.

The parent *RE*FeAsO compound is non superconducting and shows a structural distortion from the tetragonal *P4/nmm* to the orthorhombic *Cmma* space group coupled with a static antiferromagnetic long range ordering *SDW* at around 150K [1-4]. This ordering can be clearly seen in resistivity as well as heat capacity measurements. With the doping of carriers, the spin density wave behavior is suppressed along with the structural distortion. Carrier doping and occurrence of superconductivity with structural distortion is under debate. It has been argued that superconductivity occurs when magnetically ordered orthorhombic phase is suppressed and compound shows a metallic behaviour down to the superconducting transition [7]. On the other hand, through high resolution synchrotron x-ray diffraction [8] and muon spin relaxation [9] studies it has been observed that redistribution of electronic structure occurs due to structural deformation with carrier doping by F. Orthorhombic symmetry and magnetism also persists along with superconductivity in lower F doping range [8-9]. In higher doping ranges (≥ 15%) the orthorhombic symmetry and magnetism get suppressed [8-9]. Occurrence of superconductivity in oxy-pnictide compound is rather similar to cuprate in which superconductivity appears when adequate charge carriers are transferred in to the $CuO_2$ plane by chemical doping in charge reservoir layer [10-11].

Here, we report synthesis, structural detail, electrical, magneto-transport and specific heat of $SmFeAsO_{1-x}F_x$ ($x$ = 0.0, 0.2) compounds. Both the compositions are crystallized in a tetragonal structure with space group *P*4/*nmm*. Superconductivity in $SmFeAsO_{0.80}F_{0.20}$ sample is confirmed by resistivity and magnetization measurements. Upper critical field, [$H_{c2}(0)$] is estimated through $\rho(T, H)$ measurements using Ginzburg-Landau (GL) theory. Activation energy $U_0$ for *TAFF* is calculated by fitting of Arrhenius plot. *AC* susceptibility measurements are carried out in varying fields and frequencies to know the vortex dynamics. Current density $J_c$ has been estimated by inter-grain peak temperature $T_P$. Specific heat measurements of undoped SmFeAsO compound reveal that *SDW* transition occurs around 140K.

**Experimental details:**

In the current report we have studied the parent and optimally F doped samples. However, it is possible to dope F in various compositions we here opted for the composition which results in maximum $T_c$. Bulk polycrystalline SmFeAsO and $SmFeAsO_{0.8}F_{0.2}$ compounds are synthesized via solid-state reaction route using vacuum encapsulation technique [12]. Stoichiometric amounts of Sm, As, Fe, $FeF_3$, and $Fe_2O_3$ of better than 3*N* purity are mixed in ratio of $1/15(FeF_3) + (Sm) + 4/15(Fe_2O_3) + 2/5(Fe) + As$, in a glove box of <1 ppm level of oxygen and humidity. The mixed powder is pelletized in rectangular bar shape and further encapsulated in an evacuated ($10^{-3}$ Torr) quartz tube. Sealed and evacuated quartz tube is heat treated for 12 h at 500 $^0$C, 750 $^0$C, 950 $^0$C and finally at 1100 $^0$C for 30 h and then to cooled down to room temperature. The resultant compound is re-sintered at 1150 $^0$C for 12 h.

The room temperature X-ray diffraction pattern is taken by using Rigaku X-ray diffractometer with Cu $K_\alpha$ radiation. The Rietveld refinement of XRD pattern is carried out through FullProf software. The resistivity measurements were carried out by a conventional four-probe method on a quantum design Physical Property Measurement System (PPMS-14 Tesla). Magnetization, *AC* susceptibility, and specific heat measurements were also carried out on the PPMS.

**Results and discussion:**

Fig. 1 show the observed and Rietveld fitted room temperature *XRD* pattern of SmFeAsO and SmFeAsO$_{0.8}$F$_{0.2}$. The Rietveld analysis of the *XRD* pattern confirmed that both the studied samples are crystallized in the tetragonal phase with P4/*nmm* space group. The samples are nearly single phase with some weak impurity lines (observed in F-doped sample) are marked with * in *XRD* pattern. There is no un-reacted iron are observed and impurity lines are due to SmAs or FeAs and Sm$_2$O$_3$. Both samples are apparently good enough for the physical properties measurements. The lattice parameters are *a* = 3.937(2) Å, c = 8.492(1) Å and Vol. = 131.64 Å$^3$ for SmFeAsO and *a* = 3.926(2) Å, c = 8.460(4) Å and Vol. = 130.42 Å$^3$ for SmFeAsO$_{0.8}$F$_{0.2}$. The decrease in c-parameter and the volume indicative of successful substitution of F$^{-1}$(R$_F$ = 1.33 Å) at O$^{2-}$ (R$_O$ = 1.40 Å) site. All the permitted planes of the *RE*FeAsO system are shown by the blue vertical lines between the observed/fitted patterns and their difference in the bottom. These results are in agreement with earlier reports [3-5]. The Sm and As located at the Wyckoff position 2c (1/4, 1/4, *z*), Fe at 2b (3/4, 1/4, 1/2), and O/F are shared at site 2a (3/4, 1/4, 0). The value of Sm(*z*) is 0.139(6) and As(*z*) is 0.655(4) for SmFeAsO. On the other hand, Sm(*z*) is 0.142(6) and As(*z*) is 0.660(2) for SmFeAsO$_{0.8}$F$_{0.2}$.

Resistivity versus temperature ρ(T) plot for SmFeAsO and SmFeAsO$_{0.8}$F$_{0.2}$ compounds are shown in Fig. 2(a). The ρ(T) curve for the parent SmFeAsO compound exhibits sudden decrease in resistivity at around 150K. The decrease in resistivity below SDW ordering is due to increased mobility of the carriers which not participate in SDW state [13], loss of spin-disorder scattering may also be contributing. This anomaly in resistivity is due to the *SDW* instability and the structural phase transition from tetragonal to orthorhombic phase. It is in close agreement with earlier reports on similar compounds [4]. The SmFeAsO$_{0.8}$F$_{0.2}$ compound does not show any anomaly in resistivity. The normal state resistivity of SmFeAsO$_{0.8}$F$_{0.2}$ sample is decreased by a factor of three before the superconductivity sets in at *T$_c$* (onset) of around 51.7K, thus exhibiting good metallic normal state behavior. Finally the superconductivity is established at *T$_c$* (ρ = 0) of 44.8K.

Fig. 2 (b) shows the temperature dependence of the resistivity ρ(T, H) in applied magnetic field from 0 to 130kOe for SmFeAsO$_{0.8}$F$_{0.2}$ sample. The resistive transition shifts to the lower temperature by applying magnetic field. Although the transition width broadens with applied magnetic field, but the onset transition temperature is not much sensitive to magnetic field and is almost same. Here, we define transition temperature $T_c(H)$, which satisfies the condition that $\rho(T_c, H)$ equals to 90% of the normal-state value $(\rho_N)$ for applied field $H$. The $T_c(H)$ values are 51.7 K and 50.15 K for 0 and 130kOe field respectively indicating that the upper critical field is very high for this superconductor or a characteristic of type-II superconductor. The calculated value of $dH_{c2}/dT$ is thus 13.5 kOe/K. The transition width $\Delta T_c = T_c(90\%) - T_c(10\%)$ are 5.081 and 11.24K for 0 and 130kOe respectively. Inset of Fig. 2 (b) shows the temperature derivative of resistivity for the superconducting sample SmFeAsO$_{0.8}$F$_{0.2}$ at various applied magnetic field. The temperature derivative of resistivity gives narrow intense maxima centered at $T_c$ in zero applied fields confirming that the grains maintain a good percolation path between them. The resistivity peak is broadened under applied fields and diminished with increasing field as flux penetrates in the individual grains. It is also clear that, derivative of resistivity shows only one peak which is different from the Y(Ba$_{1-x}$Sr$_x$)$_2$Cu$_3$O$_{7-\delta}$ compound, in which two different peaks are seen clearly for inter and intra-grain with increase in applied field [14-17]. The broadening and separation of the peak increases with applied magnetic field and a broad peak is observed at low temperatures representing the intra-grain and inter-grain regimes [18].

The upper critical field [$H_{c2}(T)$] values at 0K are calculated by Ginzburg-Landau (GL) theory, which not only determines the $H_{c2}$ value at zero Kelvin [$H_{c2}(0)$], but also determines the temperature dependence of critical field for the whole temperature range. The $H_{c2}(T)$ is determined using different criterion of $H_{c2} = H$ at which $\rho = 90\%\rho_N$, $50\%\rho_N$ and $10\%\rho_N$, where $\rho_N$ is the normal resistivity or the resistance at about 52 K. The Ginzburg-Landau (GL) equation is:

$$H_{c2}(T) = H_{c2}(0) * [(1 - t^2)/(1 + t^2)]$$

Here, $t = T/T_c$ is the reduced temperature. The fitting of experimental data according to the above equation, $H_{c2}(10\%)$, $H_{c2}(50\%)$ and $H_{c2}(90\%)$ are estimated to be 499.17, 1038.44 and 3774.2kOe respectively at 0K. The variations of $H_{c2}(T)$ with temperature are shown in Fig. 2(c).

If we consider $H_{c2}$ at 90% of the resistive transition and we evaluate an average slope $dH_{c2}/dT = 4.8$ TK$^{-1}$, the corresponding $H_{c2}(T = 0)$ value derived from the Werthamer–Helfand–Hohenberg (WHH) formula is

$$H_{c2}(0) = -0.693 T_c dH_{c2}/dT |_{T_c} \approx 201 \text{ Tesla}$$

The value is comparable to other reports, although WHH evaluation underestimates $H_{c2}$ in two-band superconductors; very high upper critical fields are expected from these compounds.

The Arrhenius plot of resistivity under various applied field are shown in main panel of Fig. 3. The broadening of superconducting transition temperature in resistivity curve under applied magnetic field is due to the thermally activated flux flow of the vortices (*TAFF*) [19, 20]. The temperature dependence of resistivity in broaden region is given by Arrhenius equation [20]

$$\rho(T, B) = \rho_0 \exp\left[-U_0/k_B T\right]$$

Where $U_0$ is the activation energy for thermally activated flux flow, $k_B$ is the Boltzmann's constant and $\rho_0$ is the pre exponential factor. *TAFF* activation energy can be obtained from the slope of the linear part of an Arrhenius plot. The best fit of the slope of $\ln(\rho/\rho_{55})$ Vs $1/T$ plot gives value of the activation energy ranging from $U_0/k_B$ = 1215.2K to 307.5K in the magnetic field range of 0.1 T to 13T. The magnetic field versus the activation energy, $U_0$ plot is shown in inset of Fig. 4. The activation energy shows weak dependence i.e. $U_0/k_B \sim H^{-0.13}$ at lower fields but strongly decreases as $U_0/k_B \sim H^{-0.60}$ for higher field range. These values are good agreement with earlier report [21].

Fig. 4(a) depicts the *DC* magnetic susceptibility versus temperature plot in both zero-field-cooled (*zfc*) and field-cooled (*fc*) situations under 10Oe applied magnetic field. The negative susceptibility at 48K in both *zfc* and *fc* situation gives the clear evidence of the bulk superconductivity. It is interesting to note that the transition $T_c$ (onset) seen in magnetization is slightly lower than that obtained from the resistivity. In general, the difference in transition temperature, $T_c$ (onset) obtained from transport $\rho(T)$ and magnetic $M(T)$ measurement is observed in superconductors because transport measurements is through percolation and hence is less in comparison to bulk diamagnetic response. Therefore we get higher $T_c$ (onset) in transport measurements in comparison to magnetic measurements. The shielding fraction is determined above 22% and superconducting volume fraction around ~ 8% in the superconducting sample. Some other impurity or defects contributions such as positive paramagnetic contribution of Sm/Fe moments in magnetic susceptibility, the pinning defects and impurities may differ, the actual superconducting volume fraction.

Isothermal magnetization *M(H)* loops of the studied SmFeAsO$_{0.8}$F$_{0.2}$ compound at 2, 5 and 10K, with applied fields of up to 20 kOe is shown in Fig. 4(b). The wide opening of *M(H)* loops up to 20kOe in superconducting state confirms the bulk superconductivity in the studied sample. The inset of Fig. 4(b) shows the first quadrant of isothermal magnetization *M(H)* loop at 2, 5 and 10K for SmFeAsO$_{0.8}$F$_{0.2}$ sample. As we increase the applied magnetic field the magnetization first goes to negative and after certain field its invert towards positive. The field of inversion at which magnetization curve deviate or invert is known as the lower critical field. The H$_{c1}$ values are found to be 1200, 800 and 500 Oe at 2, 5 and 10 K respectively. As far as irreversibility field (H$_{irr}$) is concerned the *M(H)* loops are wide open for higher fields, well this is in agreement with reference [22]. The critical current density (*J$_c$*) of superconducting SmFeAsO$_{0.8}$F$_{0.2}$ compound (2.68 mm×1.30 mm×1.70 mm) is calculated by Bean's critical state model. The *J$_c$* at 5K and 10kOe field are estimated to be $1.6 \times 10^4$ A/cm$^2$. This is comparable to earlier reported values in for similar bulk samples [22].

It has been reported that in *RE*FeAsO superconductors the superconductivity is hampered by presence of unreacted Fe/FeO$_x$, as these oxide are known to order magnetically above or close to room temperature [23, 24]. In order to find out any un-reacted magnetic impurity we have done the magnetization of SmFeAsO$_{0.8}$F$_{0.2}$ in normal state (above superconducting transition) i.e. 50 to 250K. The normal state *M(T)* plot is shown in Fig. 4 (c). The measurements are done under 10kOe applied magnetic field. It can be seen that the studied SmFeAsO$_{0.8}$F$_{0.2}$ sample is paramagnetic in nature in normal state with absence of any ordered magnetic impurity. Reciprocal of magnetic susceptibility, $\chi^{-1}(T)$ behaviour is linear in the measured temperature range. The magnetic susceptibility '$\chi(T)$' follows Currie-Weiss paramagnetic behaviour, similar to the compound NdFeAsO$_{0.8}$F$_{0.2}$ [6]. We also carried out *M(H)* at 250 K in varying field of up to 2kOe, to further confirm for any presence of ordered Fe/FeO$_x$ phase [see inset of Fig.4 (c)]. The linearity of *M(H)* at 250 K excludes any possibility of ordered ferrous or ferrous oxide particle in studied compound.

The real ($\chi'$) and imaginary ($i\chi''$) components of ac susceptibility as function of temperature are measured of a polycrystalline superconducting SmFeAsO$_{0.8}$F$_{0.2}$ sample to probe the detail inter-grain and intra-grain vortex dynamics [25-28]. Fig. 5(a) represent the *AC* susceptibility ($\chi'$ and $i\chi''$) versus temperature behavior of superconducting sample measured at the various frequencies, ranging from 33 to 9999Hz at 2Oe *AC* drive field. It is observed that the diamagnetic onset transition temperature nearly same for all the frequencies, so that *T$_c$* does not depends on the applied frequency. In $\chi''(T)$ curve a broad peak is observed at well below the diamagnetic onset temperature for all the applied frequencies. This peak arises due to the gradual penetration of flux into the centre of intergranular regions.

Fig. 5(b) shows the temperature variation of the AC susceptibility (χ' and χ") measured at different amplitudes 1, 3, 5, 7, 9, & 11 Oe of *AC* field. It is observed that near $T_c$ region i.e. around ~48K diamagnetic onset is same, irrespective to applied field. On increasing field the diamagnetic transition becomes sharp and saturates after certain temperature. χ"(T) plot shows a two step transition due to intergranular at low temperature and intra-granular near $T_c$ [see inset Fig. 5(b)]. The intergranular peak shifting can be seen with the increase in applied field amplitude from 1 Oe to 11 Oe. Intergranular peak shifting is not very clear as the peak is not complete. In cuprates superconductors the real part, χ' of *AC* susceptibility follow two step transitions in corresponding to two peaks in imaginary part, χ''. Two different peak in χ'' arise due to inter-grain and intra-grain shielding. The peak near to $T_c$ arises, when the *AC* field penetrates to the centre of grains, and a peak at lower temperature is corresponds to complete penetration of the *AC* field in the centre of the sample [26-28].

In Fig. 6 specific heat $C_p(T)$ at zero field for SmFeAs$_{0.8}$F$_{0.2}$ and SmFeAsO are shown in temperature range from 2.2 to 200K. The specific heat, $C_p(T)$ value at 200K is 91.5 J/mole-K for SmFeAsO$_{0.8}$F$_{0.2}$ and 88 J/mole-K for SmFeAsO sample, which is slightly lesser than earlier reported values [29]. SmFeAsO shows a small hump in specific heat around 140K. Interestingly, this temperature is slightly lower than that of metallic step observed in resistivity measurement. This is in accordance with earlier reports [29]. This metallic step is attributed to the collective effect of *SDW* magnetic anomaly and structural phase transition in undoped ground state non-superconducting *RE*FeAsO system. With further lowering temperature $C_p(T)$ goes down before showing a sharp peak at around 4.9K. On the other hand SmFeAs$_{0.8}$F$_{0.2}$ compound shows only one shallow peak around 3.6K has been observed, the magnified view is shown in inset of Fig. 6. The peak is due to anti-ferromagnetic ordering of Sm$^{3+}$ spin [12]. This *AFM* ordering behavior is also seen in NdFeAsO$_{0.8}$F$_{0.2}$ and CeOFeAs at 1.8K and 3.7K which is due to *4f* electrons of rare earth element. In case of LaFeAsO, there are no *4f* electrons and thus no such specific heat anomaly is observed at low temperature [29-30]. Due to the decrease in magnetic moments of trivalent Sm, Ce and Nd the *AFM* ordering temperature also decrease as 5.4, 3.7 and 2.2K respectively.

We fitted the $C_p(T)$ curves with well known equation $C_p(T)/T = \gamma+\beta T^2$, where $\gamma$ and $\beta$ are the value of electronic density of states and approximate value of Debye temperature respectively. We did the fitting in the temperature range 12 K to 21 K, although some uncertainty will be there due to presence of impurity phase. The obtained values are $\gamma = 43.2$ mJ/mol-K$^2$ and $\beta = 0.25$ mJ/mol-K$^4$ for SmFeAsO and $\gamma = 92.69$ mJ/mol-K$^2$ and $\beta = 0.39$ mJ/mol-K$^4$ for SmFeAs$_{O.8}$F$_{0.2}$. The value of $\gamma$ and $\beta$ obtained by fitting for SmFeAsO is comparable to earlier report [12, 30].

In summary nearly single phase polycrystalline SmFeAsO and SmFeAsO$_{0.8}$F$_{0.2}$ samples were prepared by using single step solid state reaction method. The $T_c$ onset comes around 51.7K and finally superconductivity $T_c$ ($\rho$=0) of 44.8 K. The physical properties including structural details, electrical, thermal and magnetic are studied with applied magnetic field up 13 Tesla. Inter-grain and intra-grain vortex dynamics are investigated by *AC* susceptibility measurements. The upper critical field and thermally activated flux flow energy are estimated for SmFeAsO$_{0.8}$F$_{0.2}$ superconductor.

**Acknowledgement:**

Authors would like to thank Director NPL Prof. R. C. Budhani for his keen interest and encouragement for the study. Anand Pal and Shiva Kumar would like to thank CSIR-India for granting senior research fellowship.

**Figure Captions**

**Fig. 1** Observed and Rietveld fitted room temperature XRD patterns of SmFeAsO and SmFeAsO$_{0.8}$F$_{0.2}$. Weak impurity lines are marked with *.

**Fig. 2 (a)** Temperature dependence of the resistivity $\rho(T)$ of SmFeAsO and SmFeAsO$_{0.8}$F$_{0.2}$ samples are shown by blue and red line respectively.

**Fig. 2(b)** Resistivity behavior under applied magnetic field $\rho(T,H)$ up to 13 Tesla for SmFeAsO$_{0.8}$F$_{0.2}$. Inset shows the derivates of resistivity for SmFeAsO$_{0.8}$F$_{0.2}$, in transition region.

**Fig. 2(c)** Dependence of upper critical field H$_{c2}$(T) with temperature for the sample using SmFeAsO$_{0.8}$F$_{0.2}$ using Ginzburg- Landau (GL) equation for 90%, 50% and 10 % drop of of the normal state resistance.

**Fig. 3** The Arrhenius plot of resistivity at different field for SmFeAsO$_{0.80}$F$_{0.20}$ sample. The solid lines are linear fit of slop. Inset show the magnetic dependence of TAFF activation energy, U$_0$.

**Fig. 4(a)** Temperature dependence of magnetic susceptibility, *M(T)* of SmFeAsO$_{0.8}$F$_{0.2}$ sample in both *zfc* and *fc* situation under applied field of 10Oe.

**Fig. 4(b)** Isothermal magnetization loos *M(H)* at 2, 5, and at 10 K of SmFeAsO$_{0.8}$F$_{0.2}$ sample. Inset shows the first quadrant of *M(H)* at 2, 5, and 10K for the same sample.

**Fig. 4(c)** *M(T)* of SmFeAsO$_{0.8}$F$_{0.2}$ normal state sample, inset shows the *M(H)* at 250 K for the same sample.

**Fig. 5(a)** Real ($\chi'$) and imaginary ($\chi''$) components of *AC* susceptibility as a function of temperature for the various applied frequencies ranging from 33 to 9999 Hz at a fixed amplitude of 2 Oe for SmFeAsO$_{0.8}$F$_{0.2}$ sample. Arrows indicate the increasing frequency values.

**Fig. 5(b)** Real ($\chi'$) and imaginary ($\chi''$) components of the *AC* susceptibility versus temperature, measured in the SmFeAsO$_{0.8}$F$_{0.2}$ sample at the *AC* field amplitudes 1, 3, 5, 7, 9, & 11 Oe at a fixed frequency 333 Hz, arrows indicate the increasing *AC* field amplitudes. Inset presents the zoom in of the $\chi''$ components showing two peak transition.

**Fig. 6** Specific heat of SmFeAsO$_{0.8}$F$_{0.2}$ and SmFeAsO sample (main panel). In inset the specific heat at low temperature compared is shown.

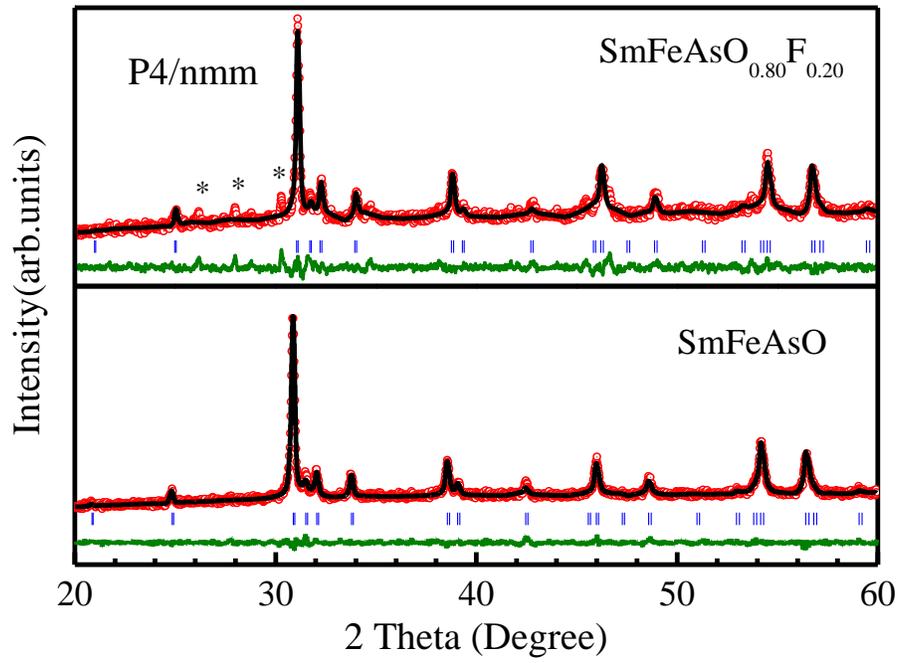

Fig.1

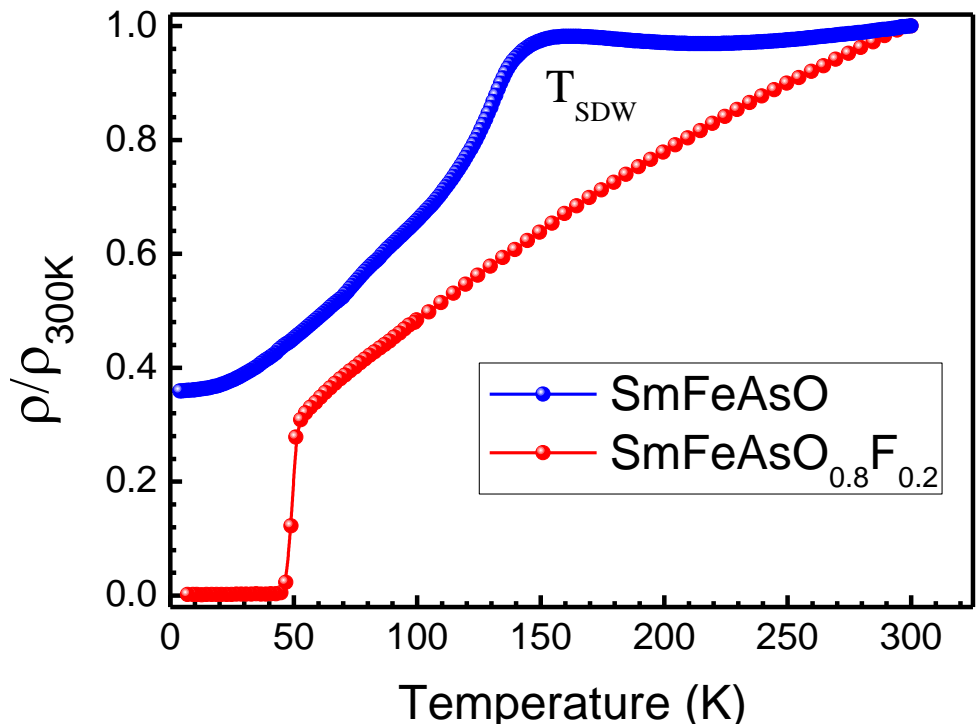

Fig.2 (a)

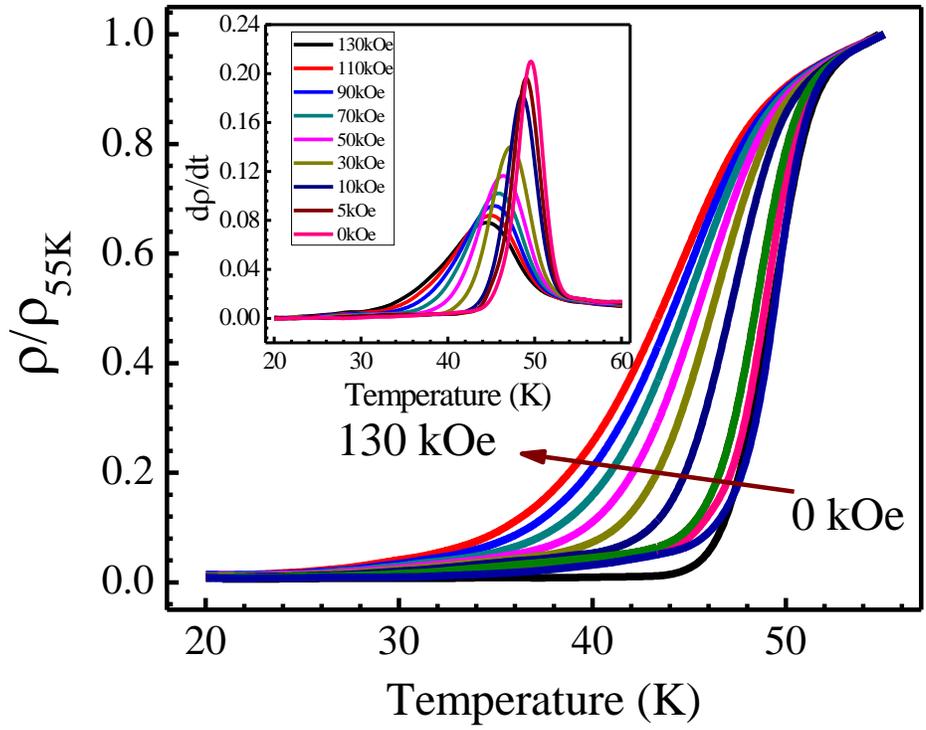

Fig. 2(b)

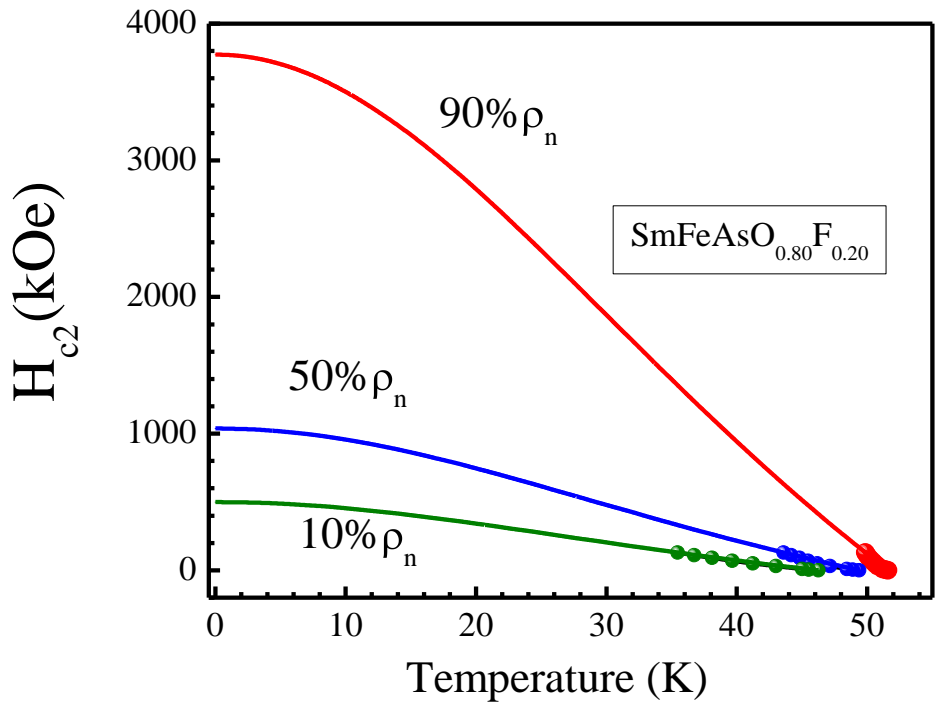

Fig.2(c)

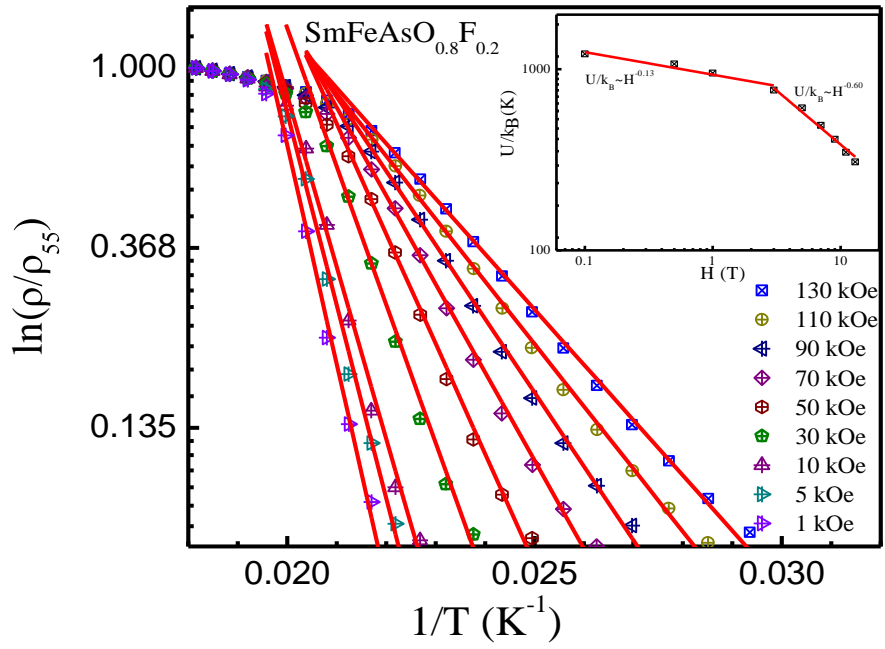

Fig. 3

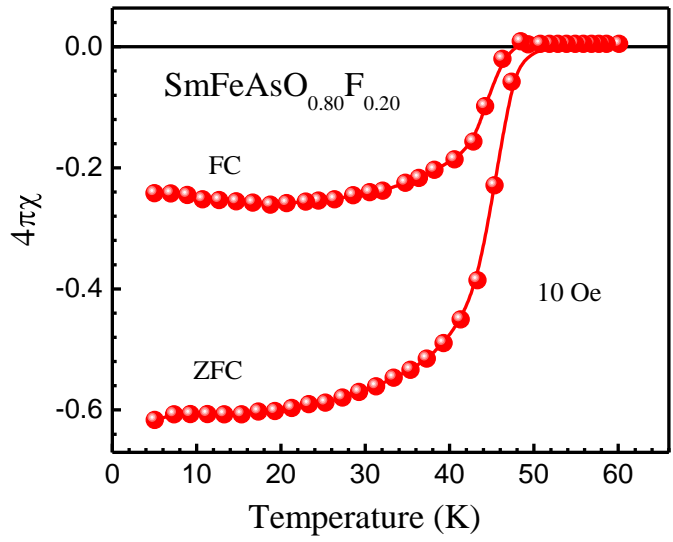

Fig.4 (a)

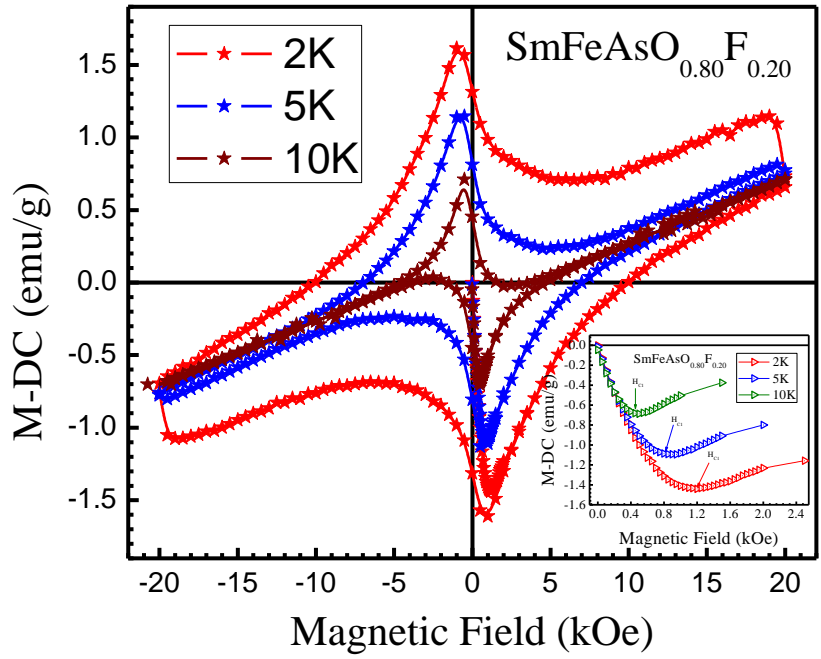

Fig.4 (b)

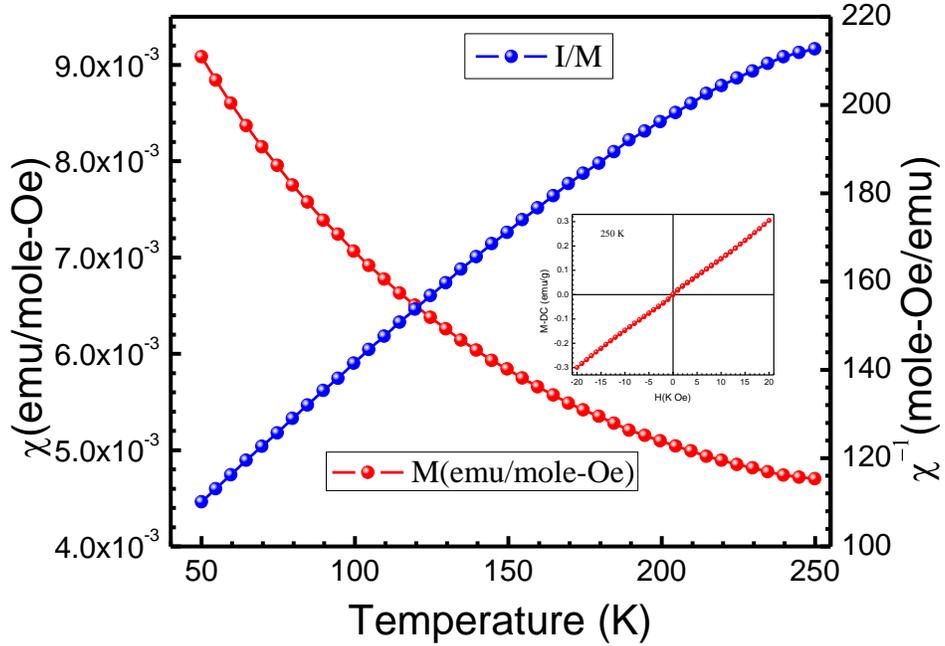

Fig.4 (c)

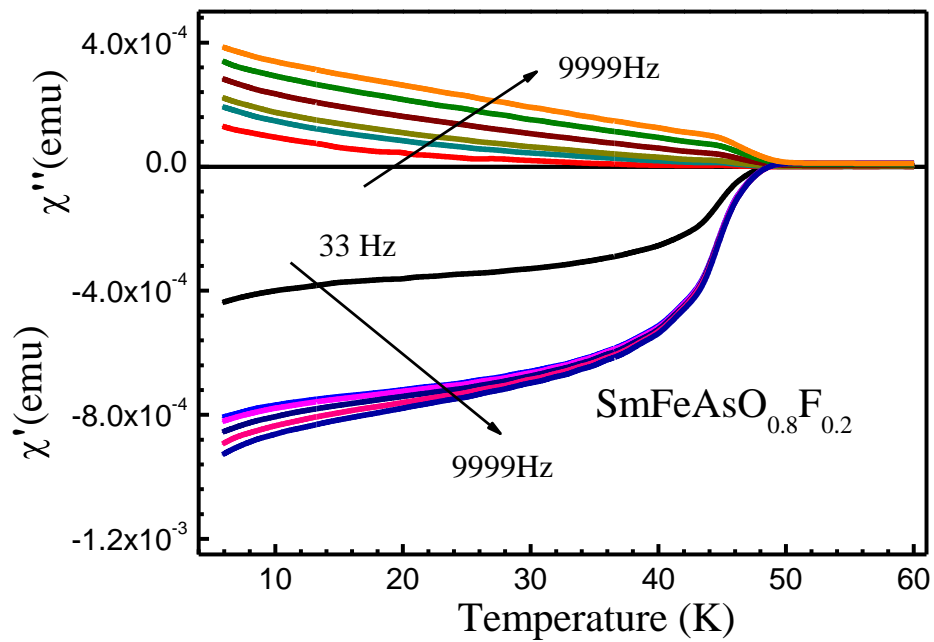

Fig. 5 (a)

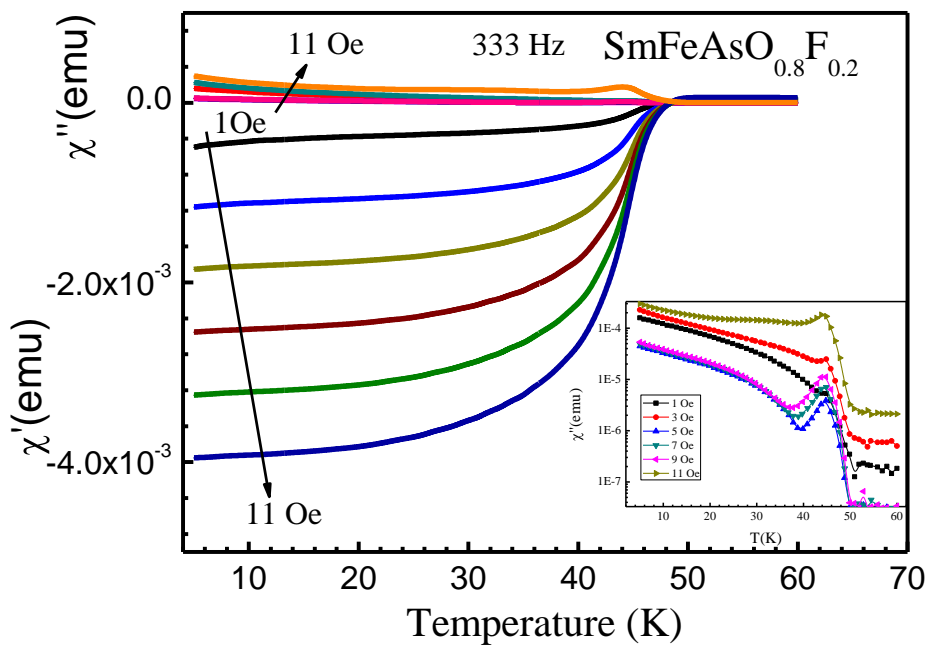

Fig. 5 (b)

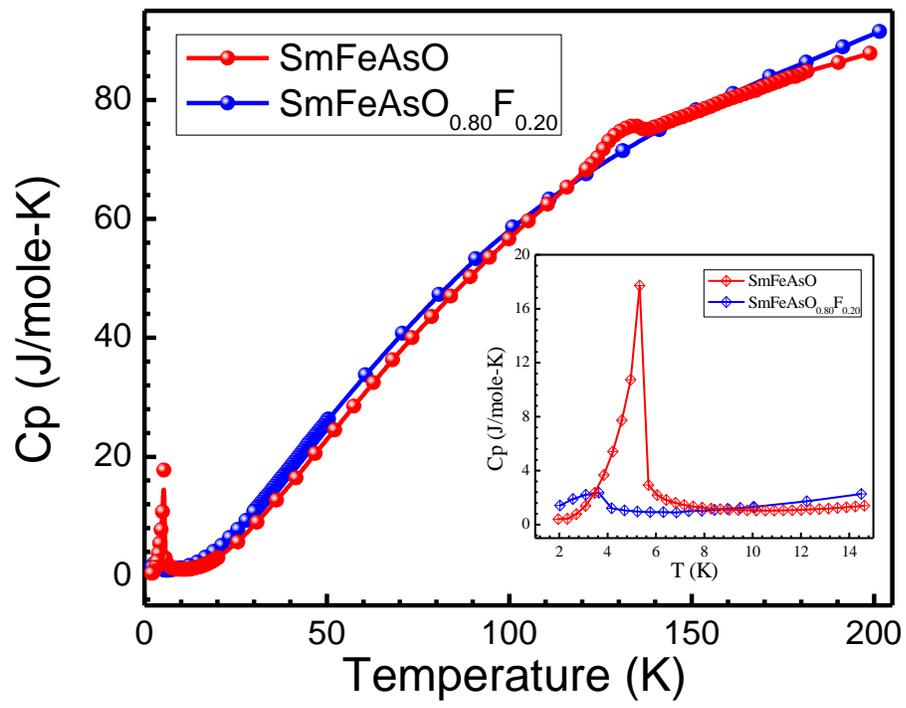

Fig. 6